\documentclass[manuscript]{acmart}
\pdfoutput=1
\AtBeginDocument{%
  \providecommand\BibTeX{{%
    \normalfont B\kern-0.5em{\scshape i\kern-0.25em b}\kern-0.8em\TeX}}}

\copyrightyear{2023}
\acmYear{2023}
\setcopyright{acmcopyright}\acmConference[]{}
\acmBooktitle{}
\acmPrice{15.00}
\acmDOI{}
\acmISBN{}

\begin{document}

\title{Towards A Visual Programming Tool to Create Deep Learning Models}

\author{Tommaso Cal{\`o}}
\orcid{0000-0002-3200-2348} 
\affiliation{%
  \institution{Politecnico di Torino}
  \streetaddress{Corso Duca degli Abruzzi, 24}
  \city{Torino}
  \postcode{10129}
  \country{Italy}}
\email{tommaso.calo@polito.it}

\author{Luigi De Russis}
\orcid{0000-0001-7647-6652}
\affiliation{%
  \institution{Politecnico di Torino}
  \streetaddress{Corso Duca degli Abruzzi, 24}
  \city{Torino}
  \postcode{10129}
  \country{Italy}}
\email{luigi.derussis@polito.it}

\renewcommand{\shortauthors}{Cal\`o and De Russis}

\begin{abstract}
Deep Learning (DL) developers come from different backgrounds, e.g., medicine, genomics, finance, and computer science. To create a DL model, they must learn and use high-level programming languages (e.g., Python), thus needing to handle related setups and solve programming errors. This paper presents DeepBlocks, a visual programming tool that allows DL developers to design, train, and evaluate models without relying on specific programming languages. DeepBlocks works by building on the typical model structure: a sequence of learnable functions whose arrangement defines the specific characteristics of the model. We derived DeepBlocks' design goals from a 5-participants formative interview, and we validated the first implementation of the tool through a typical use case. Results are promising and show that developers could visually design complex DL architectures.
\end{abstract}




\maketitle

\section{Introduction and Background}
\begin{figure*}[h!]
\centering
\includegraphics[width=0.5\textwidth]{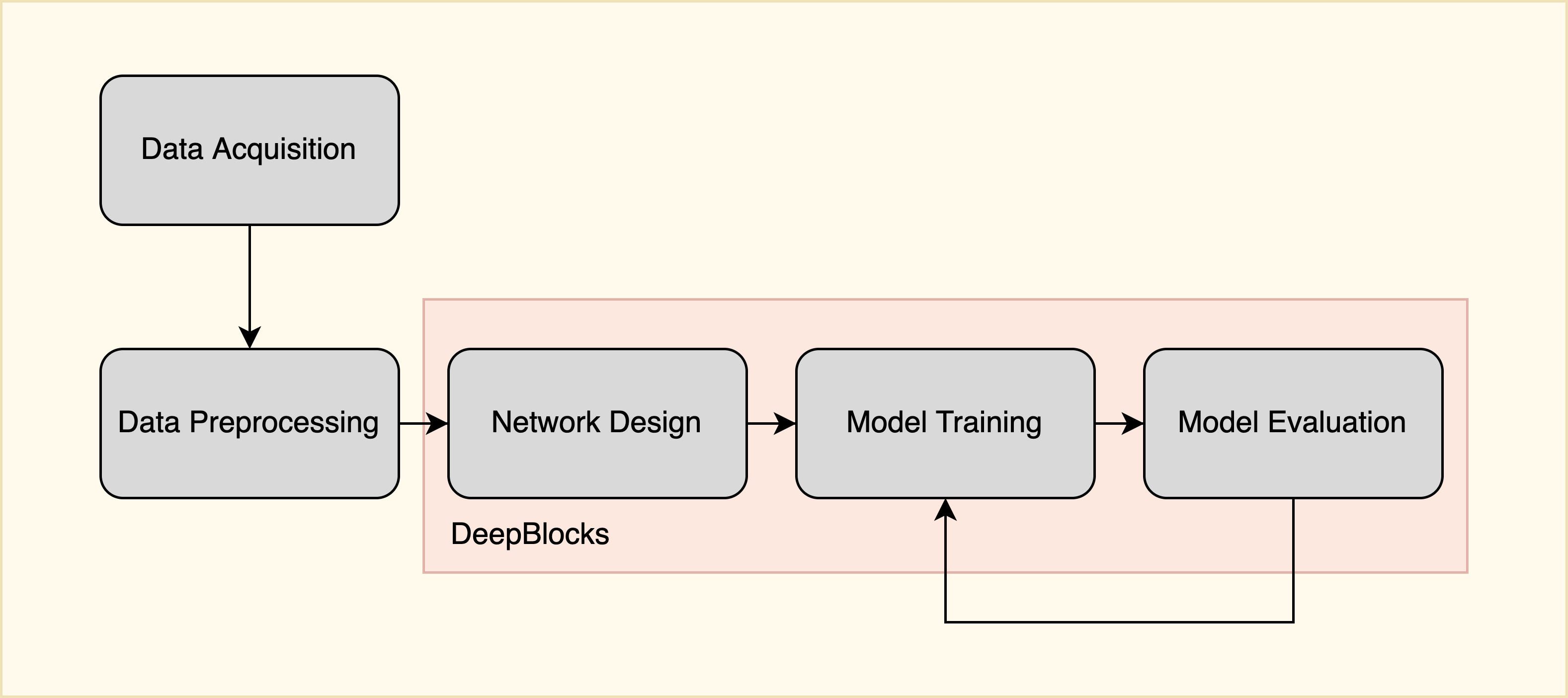}
\caption {The typical deep-learning development workflow, the highlighted part indicates the processes covered by DeepBlocks.}
\label{fig:1}
\end{figure*}
Recently, deep learning (DL) experienced rapid progress and achieved competitive performance in numerous areas such as image recognition, natural language processing, autonomous driving, medical diagnosis, and drug discovery. As such, DL developers can choose between many popular frameworks for development, including Tensorflow~\cite{tensor}, Keras~\cite{chollet2018keras}, and PyTorch~\cite{paszke2017pytorch}. These frameworks, however, require developers to have a certain level of programming skills, which many developers from diverse domains (e.g., medicine, genomics, or finance) need to master before being able to create and evaluate a DL model. Thus, many challenges arise~\cite{sank} as developers experience this steep learning curve. 
In 2017, Sankaran et al.~\cite{sank} studied the challenges that DL developers face through a quantitative survey among 113 software engineers and researchers from various backgrounds and experiences. The authors showed that DL frameworks exhibit a lack of needed features for quicker and more efficient implementation and prototyping. As a solution, 89\% suggested the need for a system able to suggest hyper-parameters and assist in debugging the DL model, while 72\% of the respondents suggested that a \textbf{visual programming tool} would be useful to speed up the overall development process. 

In traditional software development, visual programming is a paradigm that lets users create programs by manipulating elements graphically rather than by specifying them textually. DARVIZ~\cite{sank}, DL-IDE~\cite{ide},  DeepVisual~\cite{deepvis}, and ModelTracker \cite{10.1145/2702123.2702509} were the first attempts to introduce visual programming IDE enabling ``no-code'' intuitive way of designing deep learning models. They, however, exhibit limitations that do not allow the design of complex and scalable models, such as the impossibility of merging, connecting, and reusing layers and the impossibility of customizing the training procedure. In addition, they do not include important features for the complete development process, such as debugging features. Such limitations must be overcome to build larger and more complex networks, which can accomplish the recent design requirements that emerged in the community~\cite{gpt}. UMLAUT \cite{umlaut} is, instead, an example of a tool targeted to non-expert developers that focuses on debugging.

Neural Network Console developed by Sony~\cite{nnconsole} is the only available web application that overcomes most of the above limitations. However, the DL library it supports has quite a limited user scope, with less than 1\% usage over the DL community~\cite{ide}. Moreover, it is available in the cloud, only, thus limiting the possibility of using in-house machines and introducing several privacy concerns. 

To address the limitations of the available visual programming tools, and to reduce the gap between visual tools' capabilities and the freedom of expression of pure coding, we propose \textit{DeepBlocks}, a visual programming tool for DL that integrates training, debugging, and evaluation of neural network models under a single user interface. In this way, the tool covers the main steps of the DL development workflow (Fig.~\ref{fig:1}). DeepBlocks allows DL programmers to design neural networks by adding, connecting, and merging layers, which we refer to as ``blocks'', to create more complex layers and architectures. In addition, DeepBlocks allows users to create personalized blocks and add custom functions to easily adapt the tool to their application domain. 
DeepBlocks also allows developers to schedule and process multiple inputs and to design networks with multiple branches, being this a novel feature the existing tools, which only allows building layers with a single forward connection. DeepBlocks uses PyThorch, which is a DL library widely used in the DL community, adopted by up to 75\% of the papers available in the literature.

Starting from a formative interview with 3 Machine Learning engineers and 2 Ph.D. students in the field of deep learning, we obtained nine crucial functionalities to be implemented in DeepBlocks, to make it useful, efficient, and versatile. We then present the design and implementation of the tool, and report a use case to validate it. Finally, we discuss possible advantages and limitations and conclude with future work.

\section{Formative Interviews and Design Goals}
We conducted five semi-structured interviews, online and in-person, to five DL developers in October 2022. In the interviews, we focused on three main questions: 1) how they perform the design of deep learning architectures; 2) the main difficulties they face throughout the process; and 3) we discussed the possible advantages and disadvantages of the adoption of a visual programming interface to develop deep learning architectures.

We interviewed three data scientists who frequently develop deep learning models at a large, data-driven software company, and two artificial intelligence Ph.D. students, who both develop and apply deep learning models in their research field. 3 participants (P1–P5) self-identified as male and 2 self-identified as females, their age was between 25 and 29 years old, and they signed a consent form before starting the interview. We synthesized the results of the interviews into nine design goals, which guided the design of DeepBlocks. In a subsequent phase, participants were asked to rank the goals by importance and comment on their decisions.

\subsection{Results and Design Goals}
For P1, P2, and P4 the main difficulty in programming deep neural networks is to understand from the code how the network is structured; they argue that the code structure, in a large number of cases, does not reflect the structure of the network, leading to inefficient design. 

For P3 and P5 the most important issue is to locate bugs in the architecture, along with the fact that they can only spot them at the start of the training phase.
From this insight, we set our first design goal to be \textit{Interactive Debugging.}
Furthermore, P1 and P3 point out that it takes a lot of effort to monitor the inputs and outputs for every layer of the network during training and that simplify such procedure would be useful to better understand the behavior of the model. We summarize this finding in a second design goal: \textit{Visualization of blocks inputs and outputs.} Where we refer as a `block' to an elementary architectural layer. 

P4 and P5 add that the reuse of layers between different architectures is particularly prohibitive due to a lack of compatibility and difficulties in separating the individual layers from their context. From this insight, we obtain the third design goal: \textit{Load existing and custom blocks.} 

P1, P2, P4, and P5 are concerned about the possibility that a visual tool would be actually able to convey the same freedom of design that coding does; on the other hand, all participants agree that visual programming could be particularly suitable for the deep learning domain, due to the repetitiveness of layers that composes the architectures and the limited procedures schemes to train a network. Furthermore, P2, P4, and P5 argue that a simple and intuitive user interface is preferable over a more sophisticated one, even at the cost of implementing fewer customization options; this insight led us to derive a fourth design goal: \textit{Simple and Intuitive Interface.}
In addition, for P1, P2 and P5, in a visual programming interface default blocks should be organized in structured menus; from this observation, we obtain our fifth goal: \textit{Blocks Organized in structured menus.}
P1, P3, and P4 pointed out that such a visual tool should be capable of designing the same architectures that can be programmed with code, both in terms of complexity and scalability; from this point, we derive the sixth, seventh, and eighth design goals: \textit{Hierarchical Aggregation of blocks}, \textit{Customizable blocks} and \textit{Personalize optimization strategy.}
P3 and P4 evidenced that not only such a visual tool should be capable of designing the same architectures that can be programmed with code but also it should be able to do the same kind of model evaluation; this led us to set the ninth design goal: \textit{Visualization for model evaluation.}

\begin{figure*}[h!]
\centering
\includegraphics[width=0.65\textwidth]{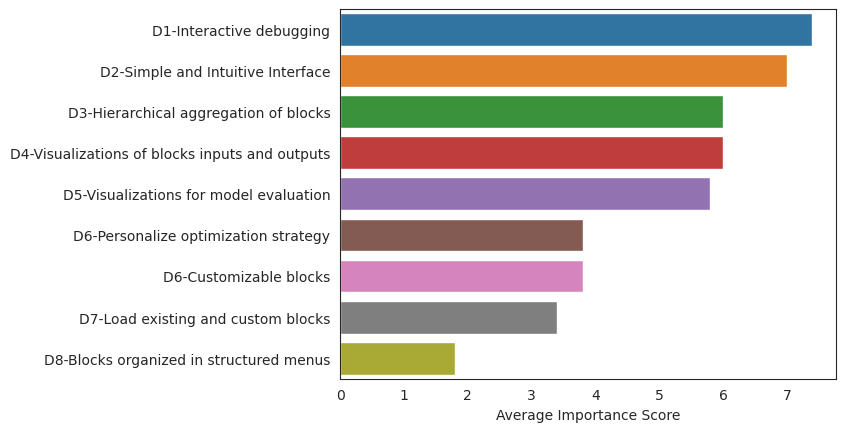}
\caption{The design goals we derived from formative interviews, ordered by the average importance given by every participant. }
\label{fig:2}
\end{figure*}

In a second phase, participants were asked to rank the resulting design goals, with a score from 1 (lowest importance) to 9 (highest importance) in order for us to focus our efforts on the most relevant design features. The results shown in Figure~\ref{fig:2} reflect the findings of the formative interviews, where participants agree to the need for a simple interface to visually program the architectures and a proper interactive debugging procedure that can easily notify users about the location and the type of bug. 

\section{DeepBlocks}
\label{sec:4}
DeepBlocks is a visual programming tool intended for Deep Learning engineers that need to develop and test complex deep learning architectures. We designed DeepBlocks to accomplish the needs extracted from the formative interviews. In this section, we list the main features of DeepBlocks and describe its design and implementation.

\subsection{An Example Scenario}
We introduce an example scenario that will be developed throughout all Section~\ref{sec:4} to illustrate to the reader how the implemented functionalities can be used by a programmer to achieve the design of a simple DL architecture. 

\begin{quote}
    \textit{John is a Computer Engineering student, he is in his third year of B.S., and he is following the Artificial Intelligence course. In the second assignment, he has been charged with developing, to classify an image dataset, a simple neural network composed of five Fully Connected Layers, using DeepBlocks.}
\end{quote}

\subsection{Layout of DeepBlocks}
\begin{quote}
    \textit{John downloads DeepBlocks installs its dependencies and launches the program. To design the network, he adds an input block and five fully connected blocks by clicking the ``+'' button on the respective voice in the right panel. By clicking on the ``Add Data\dots'' button in the ``Input'' voice on the tree menu in the left panel, John can select the dataset file from a dialog. John is not required to custom preprocess the dataset, since it is already in the format specified in the tool's documentation.}
\end{quote}

\begin{figure*}[h]
\centering
\includegraphics[width=0.9\textwidth]{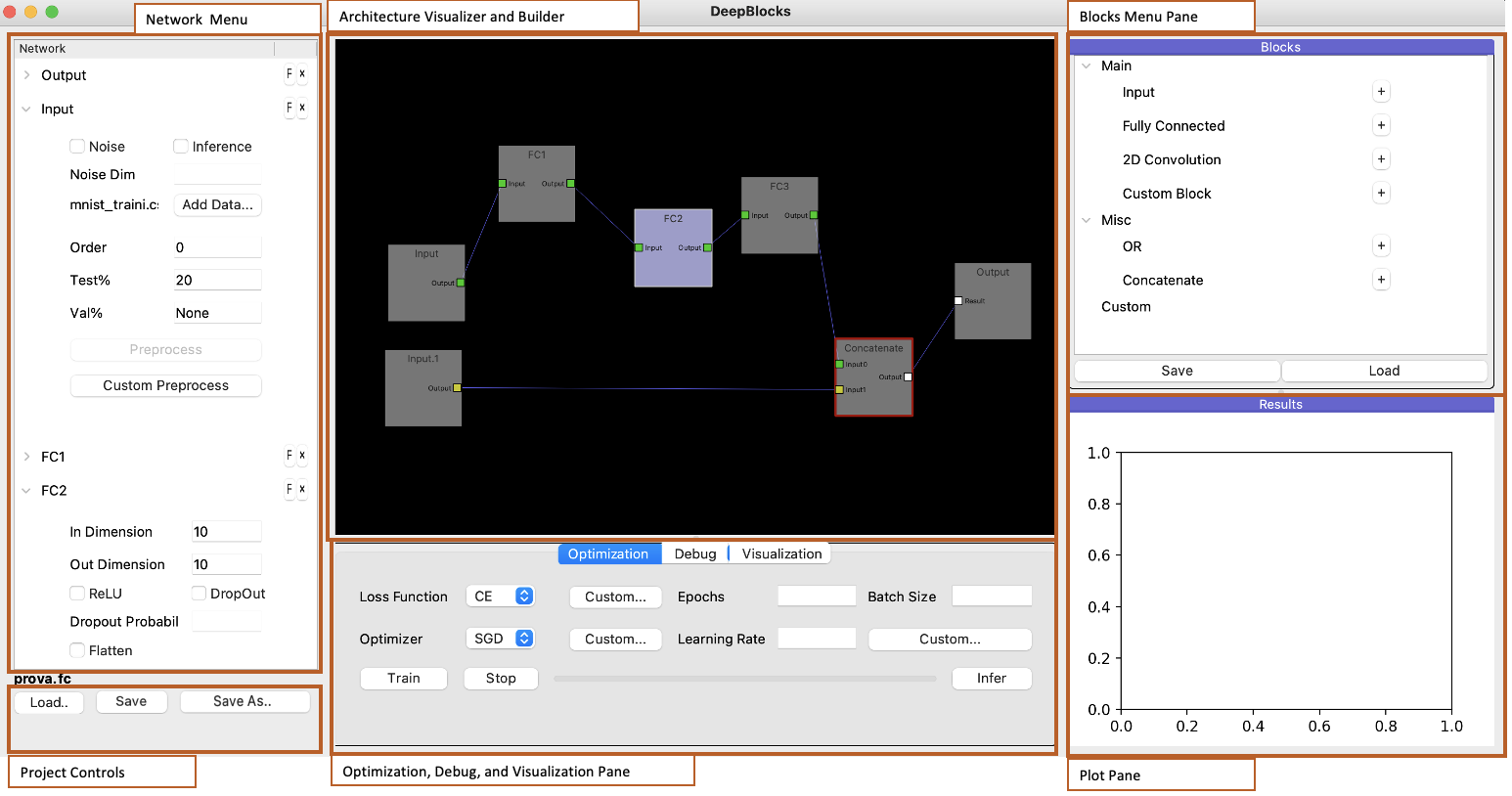}
\caption{DeepBlocks Layout}
\label{fig:4}
\end{figure*}

The layout of DeepBlocks is illustrated in Fig.~\ref{fig:4}. DeepBlocks consists of six main panes: Network Menu, The Architecture Visualizer and Builder, the Blocks Menu Pane, the Project Controls, the Results Visualization Pane, and the Optimization, Debug, and Visualization Pane. The Network Menu, located on the upper left, lists the current blocks present in the architectures and allows access to their specific controls and parameters. The Architecture Visualizer and Builder is the core of the visual programming capabilities of DeepBlocks; it visualizes the blocks and allows the user to connect the input and outputs terminal of different blocks. By right-clicking on a specific block, users can save it under the ``custom'' menu in the Blocks Pane, or, when multiple blocks are selected, users can merge them into an abstract ''SuperBlock''. In addition, the Architecture Visualizer and Builder provide debug tips: when a block is not correctly processed, its contours are colored red, and when a signal is not processed, due to an error in the block, the input and output terminals are colored yellow as shown in Fig.~\ref{fig:4}. The Blocks Pane lists the default available blocks, which are subdivided into main blocks, which are the most typical deep learning processing layers, as well as miscellaneous blocks, such as the one to concatenate data or to do a logical OR between two inputs. In addition, the Blocks Pane contains the controls that allow to saving and load one or multiple custom blocks. The Optimization, Debug, and Visualization Pane include the optimization controls to train and test the network, the debug information of the selected block, and a visualization section to visualize the inputs and outputs of every block. In the Results Visualization Pane are plotted the train and test accuracy. Finally, the Project Controls Pane, allows you to save or load a project.

\subsection{Visual Programming in DeepBlocks}
\begin{quote}
    \textit{John connects the blocks he previously added starting from the Input block, and sequentially through every Fully Connected Block until the output block. He then sets the right input and output dimension for every Block, checking the output dimension of every Block in the Visualization pane. John then merges the five fully connected Blocks into a SuperBlock and renames it ``Backbone''. He then saves the ``Backbone'' block in the custom blocks tree, by right-clicking over it and selecting the ``Save'' option. }
\end{quote}

 \begin{figure*}[h]
\centering
\includegraphics[width=.9\textwidth]{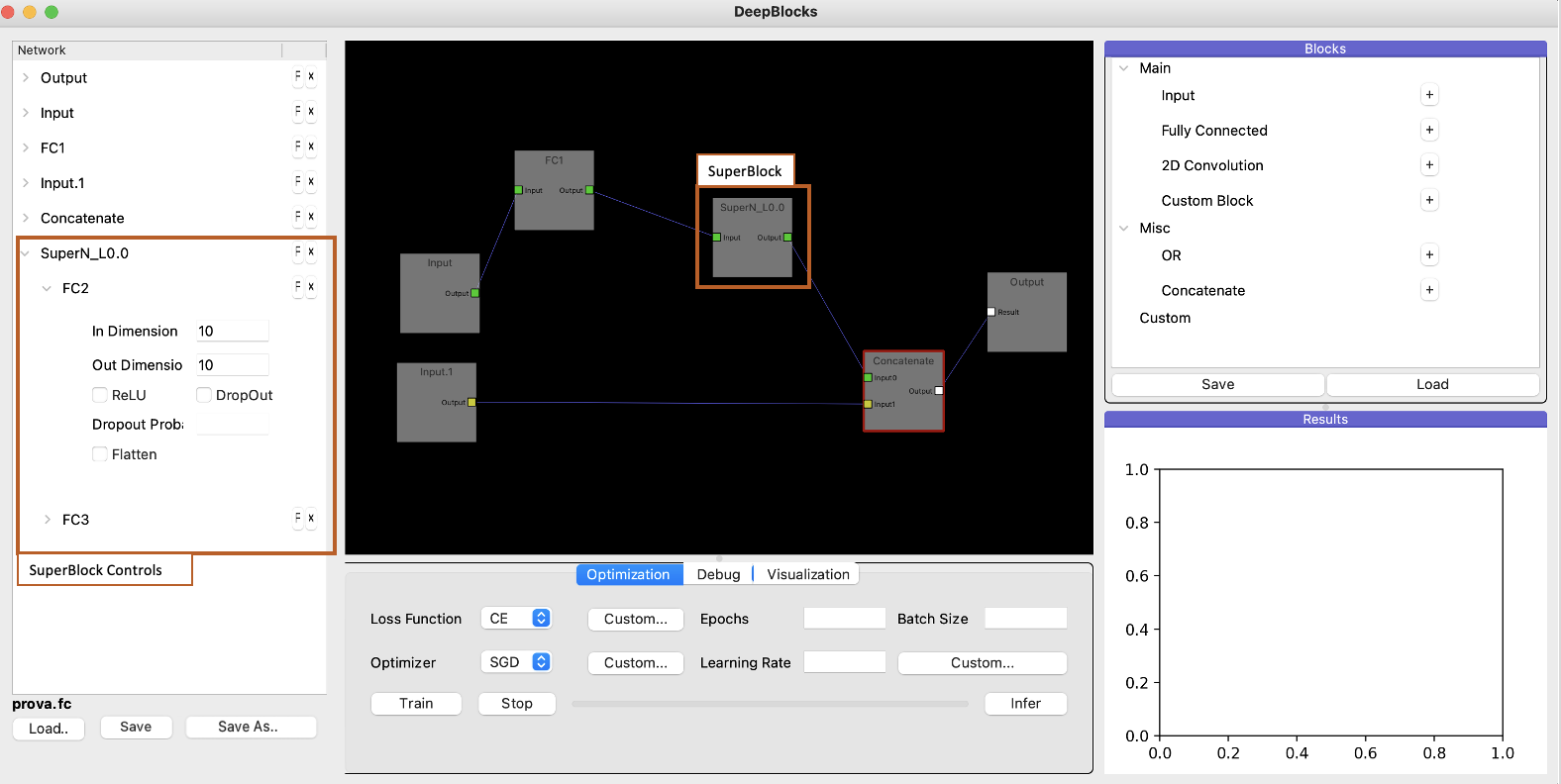}
\caption{Illustration of a SuperBlock and its hierarchical visualization in the Network Menu }
\label{fig:5}
\end{figure*}

With respect to the literature, DeepBlocks provides a fully scalable way to visually design DL architectures and the possibility to design complex, multi-branch architectures. Large practical cases could be modeled in deep blocks thanks to the possibility of adding custom blocks, merging multiple blocks into hierarchical ``SuperBlocks'' as well as the possibility of scheduling multiple inputs and creating multi-branch connections. 

A Block is composed of one or multiple input and output terminals. Every Block contains a Python function that characterizes it; users can add custom blocks by adding ``Custom Block'' from the Blocks Pane and specifying its custom function, or can directly load existing custom blocks; the latter can be useful for non-expert users, which when reusing a custom made block could only focus on its input and outputs and not on the underlying logic. The blocks specific properties can be set in the Network Menu (see Fig.~\ref{fig:5}). 

To scale up the designed architecture, selected blocks can be merged into more abstract ``SuperBlocks'' by invoking the right-click menu over the Architecture Visualizer and Builder and selecting the ``Merge'' option. SuperBlocks sub-blocks can be hierarchically visualized, along with their controls, in the Network Menu. 
To train or execute the architecture, a computational tree is generated, starting from input blocks and recurrently through every connection; if two branches converge on the same block, they are guaranteed to be processed sequentially before the successive computations. Cycles are not allowed in our setting. In order to expand the capabilities of the training procedure, we introduced a notion of ``order'' for every input. Inputs can be assigned to one or multiple orders. At every training step, orders are executed sequentially, and, for each order, only the signals coming from input blocks that belong to the specified order are passed downstream, while for the others is passed a null value. This allows the designing of many practical architectures that require alternation of different input signals.
An input signal is a dictionary composed of the input value, the ground truth, the current order, and a flag indicating if the current is a test or a training step. By accessing the dictionary, the different custom functions can be programmed depending on the order currently executing.
We believe that the above-proposed features, which are mostly missing in the literature, can make a step forward in allowing developers to design with our visual programming tool most of the variety of the typical architectural designs. 

\subsection{Debug Features and Validation}
\begin{quote}

    \textit{John notices that the contours of the ``Backbone'' SuperBlock are painted red; he inspects the debug pane to check what error is causing the block not to process, and understands that there is a problem in the input dimensions since the image must be flattened before passing it to the Fully Connected Block. He flattens the input of the first fully connected block on the left pane, and the red contour disappears. He then set up the optimization parameters, and finally trained the network. He checks the training meters by looking at the Train Metrics Plot on the right. }
    \end{quote}

\begin{figure*}[h]
\centering
\includegraphics[width=.9\textwidth]{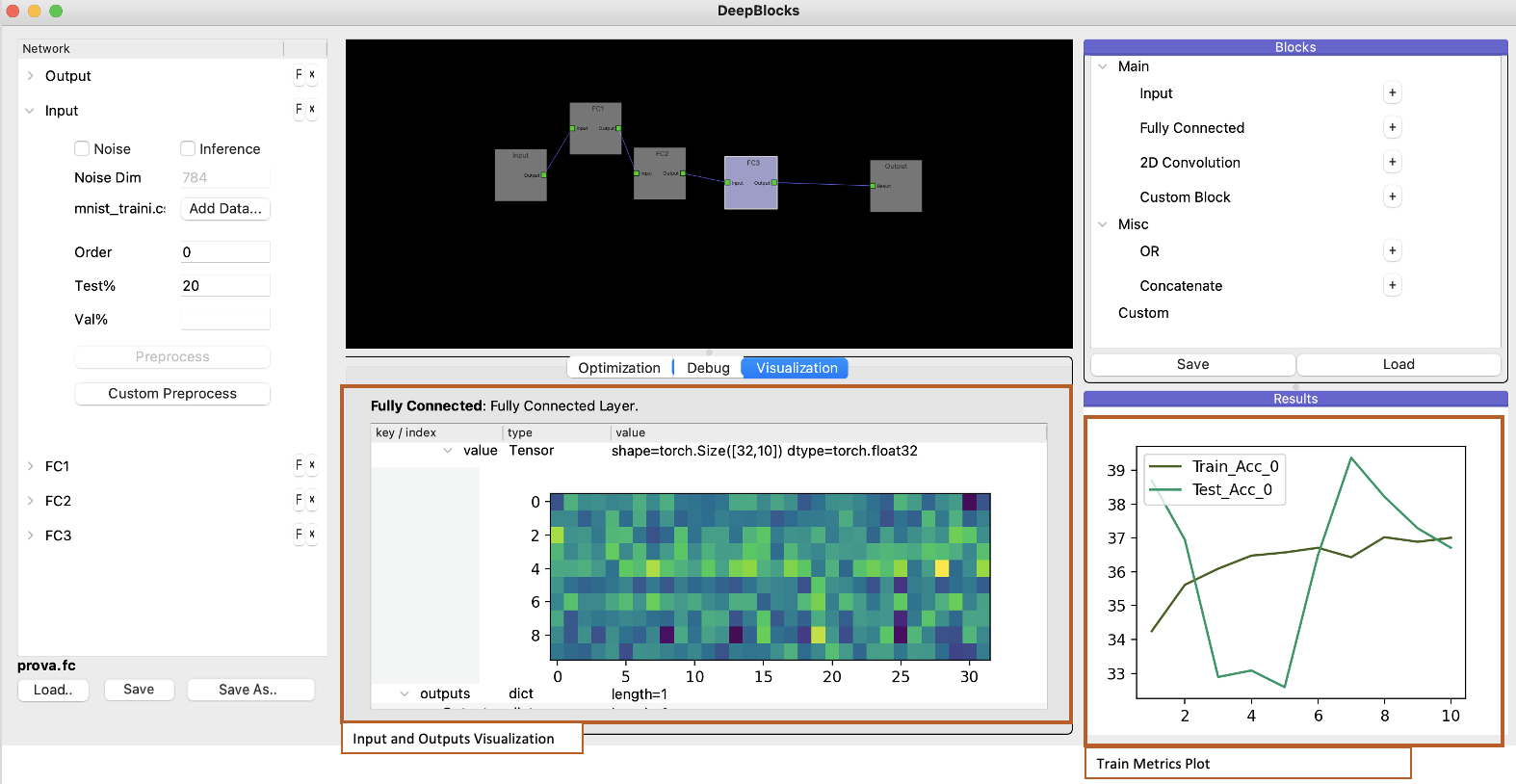}
\caption{Visualization Features of DeepBlocks. }
\label{fig:7}
\end{figure*}

Although the model evaluation is not the main focus of DeepBlocks, which is rather more engineered on the architecture design, we added two main features to monitor the model results: the Visualization and the Train Metric Plot panes (Fig.~\ref{fig:7}). The former reports, for the selected block, the input and output dimensions, as well as an heatmap of their values. The latter shows training reports evaluation metrics over the training and test data.

In addition, as in the John's story, a Debug Pane is available, showing the type and dimension of the various Blocks' inputs and outputs.

\subsection{Implementation Details}
DeepBlocks has been implemented in Python 3.8 using PyTorch \cite{paszke2017pytorch} for Deep Learning modeling, PyQT5 \cite{pyqt} for the user interface design, and PyQTgraph \cite{pyqtg} for the visual programming features.

\section{Use Case: Domain-Adversarial Neural Network}
This section demonstrates the applicability of DeepBlocks in a typical use cases: how to visually design a domain-adversarial neural network (DANN)~\cite{dann}. DANN takes as inputs labeled samples from a source distribution and unlabelled samples from a target distribution and it learns how to extract the features to solve the task for both the source and target domains.

\begin{figure*}[h]
\centering
\includegraphics[width=0.7\textwidth]{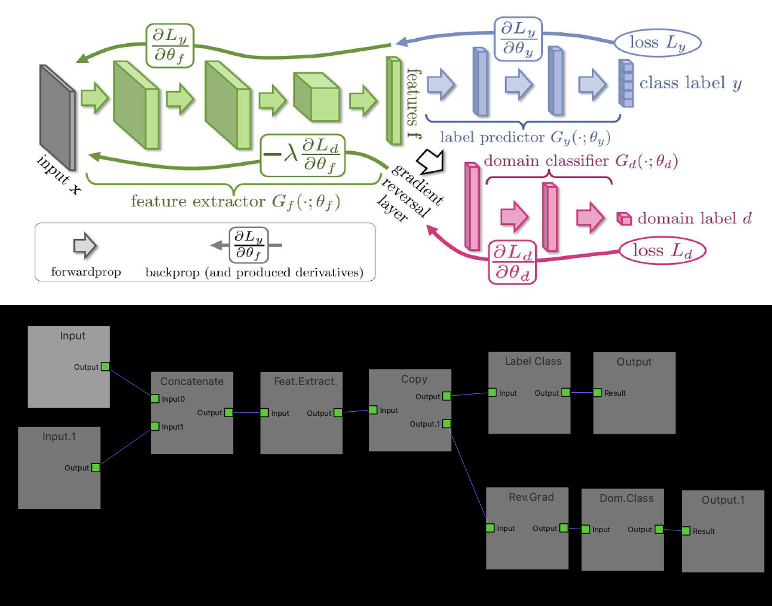}
\caption{Domain-Adversarial Neural Network as described in \cite{dann} (above) and visualized in DeepBlocks (below).}
\label{fig:8}
\end{figure*}

We start by adding two Inputs Blocks and load on the first the source and the second the target domain data. We then concatenate the outputs. To design the feature extractor, we add three Convolutional Blocks to the Architecture Visualization and Builder Pane, and connect them; we then select them and merge them into a single SuperBlock that we rename ``Feature Extractor''. From the miscellaneous blocks, we add the Copy Block that simply copies the input to one or more outputs, and we connect it to the outputs of the Feature Extractor. To design the label classifier, we add three Fully Connected Blocks, connect them and merge them as done with the Feature Extractor. We repeat the process for the Domain Classifier. To reverse the gradients of the Domain classifier, we add a Custom Block where we simply replace the predefined ``backward'' function returning its negative. Fig.~\ref{fig:8} reports the resulting architecture along with the one in the reference paper~\cite{dann}. 

\section{Conclusions and Future Work}
In this paper, we introduced DeepBlocks, a visual programming tool for deep learning software development. DeepBlocks allows developers to design and implement DL architectures visually. The tool provides several development features including model designing from scratch, interactive debugging, model training, and model inference. In addition, with respect to the previously available tools, DeepBlocks allows the designing of more complex, scalable, and custom architectures. 

We designed DeepBlocks with the support of a formative interview with 5 participants, and we preliminary validated it through a use case. With the use case, in particular, we showed that just allowing a little customization of blocks, we can permit developers to visually design complex and experimental architectures. Clearly, there is a trade-off between customization capabilities and the actual automation that DeepBlocks provides in the process of DL programming. Allowing customization without losing automation is design-challenging, but it could augment the complexity of the tool; for example, when adding the notion of ``order'' --- to let users customize the training procedure --- we require the users to specify it for every training procedure. The right balance should be found between customization capabilities, complexity, and automation of the tool. 

As future work, DeepBlocks can tame the problem of visualizing a large number of inputs and outputs (in the order of billion). Currently, there is also no way in the tool to understand the behavior of sub-blocks that composes a SuperBlock. This applies to debug as well; in fact, from a faulty SuperBlock you cannot visually locate the bug in the sub-blocks without expanding it in the Debug Pane, and this becomes unfeasible with a very large number of hierarchies. 

Among the next steps, DeepBlocks needs to undergo a series of user studies, involving its usability and effectiveness against similar tools and traditional programming approaches. Finally, once the tool is consolidated, we plan to release it and further evaluate the tool in a large-scale, in-the-wild study, e.g., with machine learning students. 

\bibliographystyle{ACM-Reference-Format}
\bibliography{deepblock.bib}

\end{document}